\documentclass[twocolumn,showpacs,preprintnumbers,amsmath,amssymb,epsf]{revtex4-1}
\usepackage{graphicx,epsfig}
\usepackage{dcolumn}
\usepackage{bm}
\usepackage{subfigure}

\def\beq{\begin{eqnarray}}
\def\eeq{\end{eqnarray}}

\begin{document}

\title{Detectable MeV neutrinos from black hole neutrino-dominated accretion flows}

\author{Tong Liu,$^{1,2}$ Bing Zhang,$^{2,3,4}$ Ye Li,$^{2}$ Ren-Yi Ma,$^{1}$ and Li Xue$^{1}$}

\affiliation{
$^1$Department of Astronomy, Xiamen University, Xiamen, Fujian 361005, China; tongliu@xmu.edu.cn; ryma@xmu.edu.cn\\
$^2$Department of Physics and Astronomy, University of Nevada, Las Vegas, NV 89154, USA; zhang@physics.unlv.edu\\
$^3$Department of Astronomy, School of Physics, Peking University, Beijing 100871, China\\
$^4$Kavli Institute of Astronomy and Astrophysics, Peking University, Beijing 100871, China}

\date{\today}

\begin{abstract}
Neutrino-dominated accretion flows (NDAFs) around rotating stellar-mass black holes (BHs) have been theorized as the central engine of relativistic jets launched in massive star core collapse events or compact star mergers. In this work, we calculate the electron neutrino/anti-neutrino spectra of NDAFs by fully taking into account the general relativistic effects, and investigate the effects of viewing angle, BH spin, and mass accretion rate on the results. We show that even though a typical NDAF has a neutrino luminosity lower than that of a typical supernova (SN), it can reach $10^{50}-10^{51}~{\rm erg~s^{-1}}$ peaking at $\sim 10$ MeV, making them potentially detectable with the upcoming sensitive MeV neutrino detectors if they are close enough to Earth. Based on the observed GRB event rate in the local universe and requiring that at least 3 neutrinos are detected to claim a detection, we estimate a detection rate up to $\sim$ (0.10-0.25) per century for GRB-related NDAFs by the Hyper-Kamiokande (Hyper-K) detector if one neglects neutrino oscillation. If one assumes that all Type Ib/c SNe have an engine-driven NDAF, the Hyper-K detection rate would be $\sim$ (1-3) per century. By considering neutrino oscillations, the detection rate may decrease by a factor of 2-3. Detecting one such event would establish the observational evidence of NDAFs in the universe.
\end{abstract}

\pacs{95.85.Ry, 97.10.Gz, 97.60.Lf, 98.70.Rz}

\maketitle

\section{Introduction}
For decades, the Sun and SN 1987A had remained as the only astrophysical neutrino sources. Recent discovery of the excess of high-energy neutrinos ($>10$ TeV) above the atmospheric background by the IceCube collaboration \cite{Icecube-nus} suggests that there exist a large population of astrophysical neutrino sources, and that neutrinos would become important cosmic messengers in the near future. In the MeV domain (the energy band where the Sun and SN 1987A were detected), multiple upcoming detectors are planned (e.g., Hyper-Kamiokande or Hyper-K \cite{HyperK}, JUNO \cite{JUNO}, and LENA \cite{LENA}). The operation of these detectors would lead to the discovery of new MeV astrophysical sources.

Whereas the leading candidates of MeV neutrino sources are supernovae (SNe) similar to SN 1987A, another type of objects, namely neutrino-dominated accretion flows (or NDAFs) around rotating black holes (BHs), have been widely discussed in the literature. These systems can be formed in massive star core collapse events or mergers of two neutron stars (NS-NS mergers) or a BH and a NS (BH-NS mergers). Such systems are believed to operate in at least some gamma-ray bursts (GRBs) \cite{GRBneutrino}, both short and long (i.e., SGRBs and LGRBs, or Type I and Type II GRBs) \cite{class}. Since NDAF neutrino emission is quasi-isotropic, NDAF systems also include those GRBs that do not beam towards Earth. Furthermore, there could be a population of core collapse events that might have an NDAF but did not produce successful relativistic jets due to their inadequate powers or lasting times \cite{failed-jets}. The NDAF models have been extensively developed over the years within the context of GRB central engine \cite{NDAF1,NDAF2,NDAF3}, but few attempts have been made to address the direct detectability of the MeV neutrinos generated by NDAFs. Assuming a distance of 10 kpc, the detectability of a nominal NDAF with a large accretion rate by Super-Kamiokande (Super-K) has been discussed \cite{NDAF4}. However, no detailed calculations of NDAF spectra by fully considering general relativistic (GR) effects and realistic parameters based on GRB observations have been presented, and no estimates of the NDAF event rate have been discussed in the literature.

\begin{figure*}
\centering
\includegraphics[angle=0,scale=1.2]{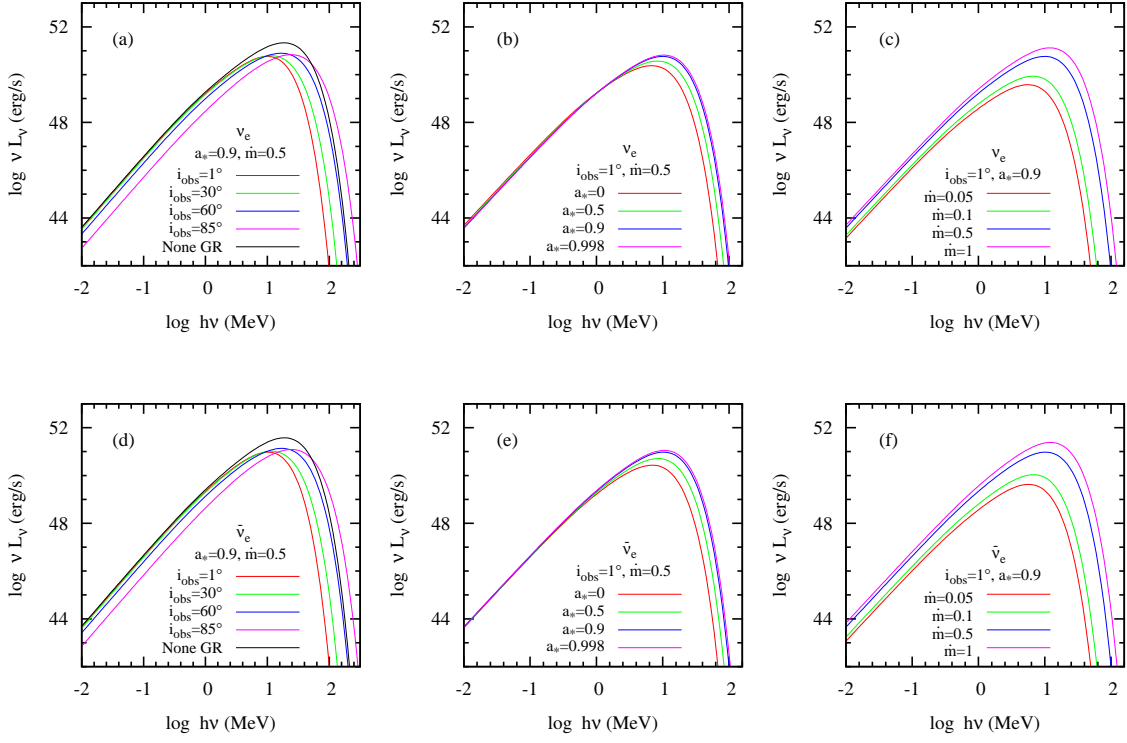}
\caption{Electron neutrino and anti-neutrino spectra as functions of viewing angle, BH spin parameter, and accretion rate.}
\label{fig1}
\end{figure*}

In this paper, we calculate the NDAF electron neutrino/anti-neutrino spectra by fully taking into account the GR effects, and investigate the effects of viewing angle, BH spin, and mass accretion rate. We will compare the predicted NDAF neutrino spectra with those of SNe, and discuss the detectability of these events by future MeV neutrino detectors such as Hyper-K.

\section{Neutrino spectra of NDAFs}
The inner region of NDAFs is extremely hot and dense with electrons in the degenerate state. Photons are trapped while neutrinos escape to carry away the viscously dissipated BH gravitational energy. The neutrino cooling processes massively occur in the disk, which is dominated by the Urca processes, but also includes electron-positron pair annihilation, plasma decay, and nucleon-nucleon bremsstrahlung. Electron neutrinos and anti-neutrinos are emitted from the surface of the disk, which are the dominant neutrino flavors. A small fraction of neutrinos and anti-neutrinos annihilate above the disk in a preferred solid angle along the axis perpendicular to the disk plane,  which would produce a thermally-dominated fireball to power GRBs. Such annihilations only consume about one percent of the total neutrino emission energy \cite{NDAF2,NDAF3}, which can be neglected when we calculate the neutrino spectra of the NDAFs. The NDAF neutrino emission is quasi-isotropic (with some angle dependences), which is detectable even if the GRB jets do not beam towards Earth.

We investigate the relativistic one-dimensional time-independent global solutions of NDAFs by taking into account detailed neutrino physics, balance of chemical potentials, photodisintegration, and nuclear statistical equilibrium \cite{NDAF3}. To avoid complications of a time-dependent problem, we consider solutions that are characterized by the mean mass accretion rate and the mean BH spin. We calculate the radial distributions of density, temperature, electron degeneracy and other physical properties in NDAFs. The electron degeneracy has an important effect on the equation of state, and the electron fraction is about 0.46 at the outer boundary of the NDAF for all solutions. Furthermore, we find that free nucleons, $\rm ^4He$, and $\rm ^{56}Fe$ dominate in the inner, middle, and outer regions, respectively. The neutrino luminosity and annihilation luminosity are calculated and the relevant analytical formulae are derived through fitting the numerical results. We notice that the neutrino trapping process can affect the value of the neutrino annihilation luminosity, especially for high accretion rate and rapid BH spin.

The neutrino cooling processes are in principle described by the Boltzmann equations. For the relativistic radial solutions of the NDAFs, a reasonable bridging formula by introducing the neutrino optical depth from scattering and absorption, shown as Equation (42) in Xue et al. (2013) \cite{NDAF3}, is introduced to replace the Boltzmann equations, which can delineate the neutrino cooling rate very well \cite{bridging}.

Based on our calculations, we derive the fitting formulae for the mean cooling rate due to electron neutrino and anti-neutrino losses, $Q_{\nu_{\rm e}}$ and $Q_{\bar{\nu}_{\rm e}}$, in units of $\rm erg~cm^{-2}~s^{-1}$, and the temperature of the disk $T$, in units of $\rm K$, as the functions of the mean BH spin parameter, mean accretion rate, and radius, which read
\beq \log Q_{\nu_{\rm e}} = 39.78 + 0.15 a_* + 1.19 \log \dot{m} - 3.46 \log r, \eeq
\beq \log Q_{\bar{\nu}_{\rm e}} = 40.02 + 0.29 a_* + 1.35 \log \dot{m} - 3.59 \log r, \eeq
\beq \log T = 11.09 + 0.10 a_* + 0.20 \log \dot{m} - 0.59 \log r, \eeq
where $a_*$ ($0 \leq a_* \leq 1$) is the mean dimensionless BH spin parameter, $\dot{m}=\dot{M}/M_\odot~\rm s^{-1}$ and $r=R/R_{\rm g}$ are the mean dimensionless mass accretion rate and radius, respectively ($\dot{M}$ and $R$ are the mean accretion rate and radius of the disk, respectively), and $R_{\rm g}=2GM_{\rm BH}/c^2$ is the Schwarzschild radius of the BH. The BH mass is not sensitive to these properties, and we adopt $M_{\rm BH}=3~M_\odot$ in our calculations. According to Equation (2), the typical energy of neutrinos from the disk is about 10 MeV. The emission rate of electron anti-neutrinos is slightly larger than that of electron neutrinos in NDAFs, because the number density of free neutrons is larger than that of free protons in the inner region of the disk, which leads to the larger production rate of electron anti-neutrinos than that of electron neutrinos in the Urca processes \cite{NDAF3}.

\begin{figure}
\centering
\includegraphics[angle=0,scale=0.5]{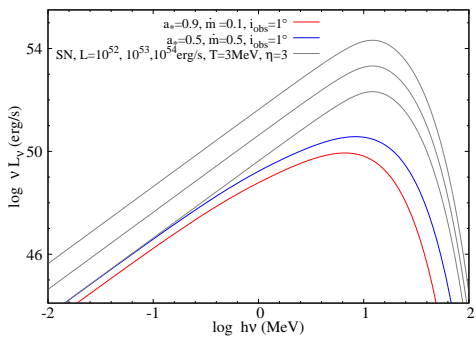}
\caption{Electron neutrino spectra of typical SGRBs (blue line), LGRBs (red line), and O-Ne-Mg core-collapse SNe (gray lines).}
\label{fig2}
\end{figure}

The neutrino cooling rate decreases with radius due to the drop of temperature and density, which suggests that neutrinos are mainly emitted from the inner region of the disk. As a result, the GR effects would be important to shape spectra. Whereas such effects were not considered by most previous NDAF calculations, here we treat these effects in full. Since the rest mass of neutrinos is much less than their kinetic energy, we treat the neutrino propagation similar to photon propagation near the accreting BH \cite{photonGR}.

We adopt the well-known ray-tracing method to calculate neutrino propagation effects \cite{GReffects}. Based on the null geodesic equation \cite{Eq}, for each pixel of the image we observe, the position of the emitter on the accretion disk can be traced numerically. The energy shift (both Doppler shift and gravitational redshift) of a neutrino can be calculated by taking into account the corresponding velocity and gravitational potential of the emission locations. Integrating over all the pixels, the energy extension of a particular rest-frame neutrino emission energy (or effectively the ``line'' profile) can be derived. Integrating over all the emission energies, the total observed spectrum is obtained. Specifically, the total observed flux can be expressed as
\beq L_{E_{\rm obs}} &=& \int dE_{\rm em} \int_{\rm image} \epsilon(R,~E_{\rm em})g^4 \nonumber \\&& \times ~\delta(E_{\rm obs}-g E_{\rm em}) ~d\Xi, \eeq
where $E_{\rm obs}$ is the observed neutrino energy, $E_{\rm em}$ is the neutrino emission energy from the local disk, $\Xi$ is the solid angle of the disk image to the observer, $g \equiv E_{\rm obs}/E_{\rm em}$ is the energy shift factor, and $\epsilon$ is the local emissivity \cite{GReffects}.

The emissivity is calculated according to the cooling rate $Q_\nu$ (either $Q_{\nu_{\rm e}}$ or $Q_{\bar{\nu}_{\rm e}}$) as
\beq \epsilon(R, E_{\rm em})=Q_{\nu} \frac{F_{E_{\rm em}}}{\int F_{E_{\rm em}} dE_{\rm em}}, \eeq
where $F_{E_{\rm em}} = E_{\rm em}^2/[{\rm exp}(E_{\rm em}/kT-\eta)+1]$ is the un-normalized Fermi-Dirac spectrum.

For a given pixel of the image, the neutrino trajectories from the emitter must satisfy the geodesic equation, i.e.  \cite{GReffects}
\beq \pm \int^\infty_{R_{\rm em}}\frac{dR}{\sqrt{l(R)}}=\pm\int^{i_{\rm obs}}_{i_{\rm em}}\frac{di}{\sqrt{I(i)}}, \eeq

For simplification some assumptions have been adopted in our calculations. First, we assume that the neutrinos are isotropically emitted from the equatorial plane, i.e., $i_{\rm em}=\pi/2$; Second, the velocity of the disk rotation is Keplerian; Third, the shading effect due to the thickness of NDAF is ignored.

\begin{table*}
\caption{The LGRB-related NDAF rate of the major galaxies in the Local Group}
\begin{tabular}{ccccccc}
\hline
\hline
Name &~~~~&Distance&~~~~& Rate of HL-GRB-related NDAF &~~~~& Rate of LL-GRB-related NDAF \\
     &    &(Mpc)   &    & ($10^{-5}$ per year) && ($10^{-5}$ per year) \\
\hline
Milky Way  && 0.01       && 7.906   &&  21.742\\
LMC       && 0.05       && 13.466  &&  37.032\\
SMC       && 0.06       && 4.212   &&  11.583\\
NGC 6822  && 0.50       && 0.378   &&  1.040\\
IC 0010   && 0.66       && 0.944   &&  2.596\\
IC 1613   && 0.73       && 0.436   &&  1.199\\
M 31      && 0.77       && 41.381  &&  113.798\\
M 33      && 0.85       && 9.908   &&  27.247\\
WLM       && 0.97       && 0.210   &&  0.578\\
\hline
\end{tabular}
\end{table*}

The effects of viewing angle, BH spin, and accretion rate on electron neutrino and anti-neutrino spectra of NDAFs are investigated in Figure 1 [(a-c) for electron neutrino, (d-f) for electron anti-neutrino]. We notice that the neutrino energies are generally in the range of 10 keV to 100 MeV, with the peak around 10-20 MeV. The spectra labeled $i_{\rm obs}=1^\circ$ may correspond to the case of GRBs, and those labeled with larger $i_{\rm obs}$ values correspond to off-beam GRBs, which could be still luminous neutrino transients. High-energy neutrinos are mainly produced from the inner region of the disk, which indicate that they would be more affected by the GR effects. As a result, as the viewing angle increases, the luminosity in the high-energy also increases (Figure 1a). Conversely, low-energy neutrinos are mostly emitted from the outer region of the disk, so that the low energy range of the spectra show little change between the calculations with and without GR effects. In Figures 1 (b) and (c), the increasing spin parameter and accretion rate can enhance the temperature of the inner region, and hence, the neutrino luminosity, especially at high energies. The similar results for electron anti-neutrino spectra are represented in Figures 1 (d-f). In general, for typical parameters, the NDAFs have neutrino luminosities reaching $10^{50}-10^{51}~{\rm erg~s^{-1}}$, which are potentially detectable from nearby galaxies.

\section{A comparison with SNe}
Since core-collapse SNe are the dominant sources of MeV neutrino emission, it would be interesting to compare our results with SN neutrinos. As an example, we calculate the neutrino spectrum of an O-Ne-Mg core-collapse SN produced at the end of the evolution of a $\sim 8-10~M_\odot$ star. The electron neutrinos are emitted with an isotropic luminosity $\sim 10^{52} - 10^{54}~ \rm erg~s^{-1}$ \cite{SNneutrino}. The normalized spectra conform with the Fermi-Dirac energy distribution. We apply Equation (1) in Ref. \cite{SNneutrino} for the calculation, with $\eta=3$ and $kT=3$ MeV adopted. For NDAFs, we adopt the typical parameters to interpret both LGRBs and SGRBs, respectively, with ($\dot{m}$, $a_*$) = (0.5, 0.5) for SGRBs, and (0.1, 0.9) for LGRBs, respectively \cite{NDAF2,lsgrbs}.

The results are shown in Figure 2. The shape of the spectra and peak energy are similar for the SNe and NDAFs. Due to the GR effect, the NDAF spectrum has a shallower slope below the peak. The main difference is luminosity. A typical NDAF neutrino luminosity is 1-4 orders of magnitude lower than a SN. In terms of duration, a SN neutrino burst may last several to 10 s of seconds. The durations of NDAFs should be consistent with the durations of GRBs, ranging from 10 s to 100 s of seconds for collapsar events and $<2$ seconds for merger events. Observations show that long GRBs are associated with Type Ic SNe \cite{GRB-SN}. It is then possible that the NDAF neutrino signals might be outshone by the SN neutrino signals. In the standard collapsar model of long GRBs that invoke the core collapse of a rapidly rotating $\sim 15 M_\odot$ Wolf-Rayet star \cite{simulation}, a proto-neutron star may survive up to seconds before collapsing into a black hole, and a GRB may be powered by accretion into the newly formed black hole with an accretion rate of about 0.1 $M_\odot~\rm s^{-1}$, lasting 10s of seconds. In these scenarios, one would expect a brief, high-flux neutrino signal followed by an extended neutrino signal with reduced flux lasting 10s of seconds. The detection of an extended low-level neutrino emission following a brief SN-related high-flux neutrino signal would be a smoking-gun signature of an NDAF.

\begin{table*}
\begin{center}
\caption{The estimated NDAF detection rates for three neutrino observatories}
\begin{tabular}{ccccccc}
\hline
\hline
Observatory &~~~~&Detection rate of GRB-related NDAFs&~~~~&Optimistic detection rate of NDAFs\\
            &    & (per century)                     &    &   (per century)\\
\hline
Hyper-K     &    & $\sim$ 0.10-0.25 (0.03-0.08)            &    & $\sim$ 1.0-3.0  (0.3-1.0)\\
JUNO        &    & $\sim$ 0.095  (0.03)                    &    & $\sim$ 0.6  (0.2)\\
LENA        &    & $\sim$ 0.095  (0.03)                    &    & $\sim$ 1.0  (0.3)\\
\hline
\end{tabular}
\end{center}

\tiny{\emph{Notes}: The numbers outside the parentheses are the detection rates without considering oscillation by requiring detecting 3 neutrinos, or the detection rates with the consideration of oscillation but requiring detecting 2 neutrinos.

The number within the parentheses are the detection rates with the consideration of oscillation and requiring detecting 3 neutrinos.}
\end{table*}

\section{Detectability}
The detectability of NDAF neutrinos by upcoming MeV neutrino detectors may be estimated in reference of SN 1987A from the Large Magellanic Cloud. Eleven events were detected by Kamiokande-II for this event, whose distance, duration, and total neutrino energy are about 50 kpc, 13 s, and $10^{53}~\rm ergs$, respectively \cite{detection}. The detectability of the MeV neutrinos in the core-collapse SNe have been discussed \cite{detection,ccsn}. The sensitivity of various detectors is defined by the number of the detected inverse-$\beta$-decay events (i.e., the electron anti-neutrino is the dominant detectable flavor) for a SN with the total neutrino energy $\sim 3 \times 10^{53}$ ergs located at 10 kpc. This number is estimated as about 170,000-260,000, 5,000, and 15,000 for Hyper-K, JUNO, and LENA, respectively \cite{HyperK,JUNO,LENA}.

The total energy of electron anti-neutrinos of an NDAF $E_{\bar{\nu}_{\rm e}}$ is adopted as about $10^{52}$ ergs. If one defines a ``detection'' as detecting three neutrino events from a certain direction within a short duration of time, one may estimate the radius of a ``horizon'' within which an NDAF is detectable by various detectors. This is $\sim$ 0.61-0.77 Mpc, 0.10 Mpc, and 0.18 Mpc for Hyper-K, JUNO and LENA, respectively. For such a small volume, the galaxies in the Local Group has to be investigated individually. Following Ref. \cite{method}, we estimate the LGRB rate of each nearby galaxy with the corrections to the star formation rate (SFR) and metallicity applied. Since NDAFs can be observed off axis from GRB jets, a beaming correction factor $f$ is introduced. For high-luminosity LGRBs (HL-GRBs), one has $f \sim 500$, which gives an event rate density of 1.6 Gpc$^{-3}$ yr$^{-1}$ \cite{GRBrate}. For low-luminosity LGRBs (LL-GRBs) which are mildly collimated, we adopt $f \sim 5$, which gives a rate 440 Gpc$^{-3}$ yr$^{-1}$ \cite{GRBrate}. The distance, mass, SFR and metallicity of each galaxy are obtained from Ref. \cite{distance}. For galaxies without a direct metallicity measurement, we apply the correlation among mass, SFR and metallicity to estimated the metallicity \cite{metallicity}. Table I shows the HL- and LL-GRB-related NDAF rates of the major galaxies in the Local Group. The contributions from other satellite galaxies is $\ll$ 10$^{-8}$ per year, which is small enough to be ignored. The contribution of SGRB-related NDAFs is also small compared with LGRBs, so we only focus on LGRB-related NDAFs. We estimate the NDAF detection rates of Hyper-K, JUNO, and LENA as 0.10-0.25, 0.095, and 0.095 per century, respectively. It is possible that a lot more massive star core collapse events may have failed jets inside the star without emerging as successful jets. Some systems may have an NDAF around a NS rather than a BH \cite{NSNDAF}. The dominant neutrino emission from these system would be from the NS rather than from the disk. Even so, the NS cooling timescale may be shorter than the accretion timescale for some cases, so that the neutrino flux would show a rapid decrease by one order of magnitude and settle down to a level defined by the NDAF emission. This signal, if detected, may be a signature of NDAFs also. In the Local Group, if one takes the event rate of SN Ib/c \cite{SNrate} as an optimistic rate for NDAFs, then the expected detection rate for NDAFs is about (1-3) per century for the Hyper-K detector and $\sim$ (0.5-1) for JUNO and LENA (Table II).

There are about 10 SN events per second in the Universe. The integrated neutrino emission from all the core-collapse SNe in the universe forms a diffuse SN neutrino background (DSNB), which may be the dominant cosmic neutrino radiation background, and is expected to be isotropic and stationary. This background was theoretically predicted before the observation of SN 1987A \cite{background}, but has not been discovered until now. Such a background could be readily differentiated from the NDAF signal discussed here if more than one neutrinos are discovered within a short time window (say, 10s of seconds). Take JUNO as an example, the prospective event rate of DSNB is about 1.5 to 2.9 events per year all sky \cite{JUNO}. The chance to have 2 neutrinos detected within several square degrees and 10s of seconds may be estimated as $\sim 10^{-8}$. This would make an effective event rate of detecting 2 clustered neutrinos to $\sim (1.5-2.9) \times 10^{-6}$ per century, which is much smaller than the corresponding JUNO event rate numbers listed in Table II. As a result, one would be able to claim a detection of NDAF if 2 (or more ideally 3) neutrinos are discovered to be spatially and temporarily clustered. The case is stronger if a bright SN signal preceding the NDAF signal is detected.

\section{Conclusions and discussion}
We have investigated the electron neutrino and anti-neutrino spectra of NDAF systems in detail. We show that even though their luminosities are lower than that of a typical SN and their event rate is much lower than SNe, future large MeV neutrino detectors may be able to detect these events. Defining a detection as at least 3 neutrinos, Hyper-K would have a detection rate $\sim$ (0.10-0.23) per century for GRB-related NDAFs, and may detect $\sim$ (1-3) per century if one assumes that all Type Ib/c SNe have an engine-driven NDAF. Detecting one such a system would firmly establish the observational evidence of NDAFs, which so far have been only theorized.

In our above calculations for NDAFs, we did not consider neutrino oscillations both inside the massive star stellar envelope and in vacuum during the propagation from the source to Earth. In SN models, the influence of flavor transformations on the flux of electron neutrinos is more significant than that on the flux of electron anti-neutrinos \cite{oscillation}. But in vacuum, this influence on the flux of electron neutrinos is similar to that on electron anti-neutrinos. Considering the similar physical conditions and flavor distributions of NDAFs with SNe, these effects may modify the neutrino spectrum moderately and reduce the neutrino flux of NDAFs by at most a factor of 2-3 \cite{SNneutrino,detection,oscillation}. In Table II, the larger numbers outside the parentheses are the detection rates without considering oscillation by requiring detecting 3 neutrinos, or the detection rates with the consideration of oscillation but requiring detecting 2 neutrinos. The smaller numbers within the parentheses are the detection rates with the consideration of oscillation and requiring detecting 3 neutrinos.

Another method to validate the existence of NDAFs may be through detections of gravitational wave signals from NDAFs due to precession and anisotropic neutrino emission in the disk, which would be detectable by the Advanced LIGO, DECIGO/BBO, ultimate-DECIGO and LISA if the sources are in the Local Group \cite{GW}. Neutrinos and gravitational waves would serve as the messengers to probe the invisible central engine of cosmic explosions.

{\em Acknowledgments. ---}
We thank Jun Cao, Shun Zhou, Bing Jiang, Shu Luo, Kate Scholberg, and Alessandro Mirizzi for beneficial discussion. This work was supported by the National Basic Research Program of China (973 Program) under grant 2014CB845800, the National Natural Science Foundation of China under grants 11233006, 11333004, 11373002, 11473022, U1331101, and U1531130. TL acknowledges financial support from China Scholarship Council to work at UNLV.

\end{document}